\documentclass[aps,prl,twocolumn,superscriptaddress,showpacs]{revtex4}
\usepackage[normalem]{ulem}
\usepackage{float}
\usepackage[usenames]{color}
\usepackage{graphicx}
\usepackage{ulem}
\usepackage{amsfonts}
\usepackage{amsmath}
\newcommand{\zb}{{z_{\rm b}}}
\newcommand{\zcf}{{z_{\mbox{\rm\tiny CF}}}}

\begin{document}

\title{Length-Controlled Elasticity in 3D Fiber Networks}

\author{C. P. Broedersz}
\affiliation{Department of Physics and Astronomy, Vrije Universiteit, Amsterdam, The Netherlands}
\affiliation{Kavli Institute for Theoretical Physics, University of California, Santa Barbara, California 93106, USA}
\author{M. Sheinman}
\affiliation{Department of Physics and Astronomy, Vrije Universiteit, Amsterdam, The Netherlands}
\affiliation{Kavli Institute for Theoretical Physics, University of California, Santa Barbara, California 93106, USA}
\author{F. C. MacKintosh}
\affiliation{Department of Physics and Astronomy, Vrije Universiteit, Amsterdam, The Netherlands}
\affiliation{Kavli Institute for Theoretical Physics, University of California, Santa Barbara, California 93106, USA}

\date{\today}

\begin{abstract}
We present a model for disordered 3D fiber networks to study their linear and nonlinear elasticity over a wide range of network densities and fiber lengths. In contrast to previous 2D models, these 3D networks with binary cross-links are under-constrained with respect to fiber stretching elasticity, suggesting that bending may dominate their response. We find that such networks exhibit a fiber length-controlled bending regime and a crossover to a stretch-dominated regime for lengths beyond a characteristic scale that depends on the fiber's elastic properties. Finally, by extending the model to the nonlinear regime, we show that these networks become intrinsically nonlinear with a vanishing linear response regime in the limit of floppy or long filaments.
\end{abstract}

\maketitle
\noindent
Materials ranging from paper and textiles to the structural components of living cells and tissues\cite{Fletcher2010} all exhibit networks of fibers or stiff polymers. 
Such networks have extraordinary mechanical properties\cite{Janmey1990,Bausch2006,Janmey2007}. 
Their elasticity depends in part on their connectivity\cite{Maxwell1864,Broedersz2011}, in analogy with jammed matter\cite{Wyart2008a,LiuNag1998} and random network glasses\cite{Thorpe1983}. 
The mechanics of the constituent fibers, and specifically their bending rigidity can also strongly impact network elasticity. 
However, the relative importance of fiber stretching versus bending is not understood, especially in 3D.
Prior work has mostly focused on 2D networks\cite{Head2003,Wilhelm2003,Heussinger2006,Onck2005,Das2007,Missel2010} since simulations in 3D have been proven to be challenging and have usually been limited to small system size\cite{Lieleg2007,Vader,Huisman2011}. 
Significant qualitative differences are expected between 2D and 3D networks: for the typical case of binary fiber interactions, the high-molecular weight limit in 2D actually corresponds to the Maxwell central-force (CF) isostatic threshold, where stretching interactions begin to completely constrain network deformations. 
In contrast, 3D networks with binary interactions remain well below CF isostaticity. 
Thus, owing to their marginal stability, real 3D fiber networks are expected to be fundamentally more bending-dominated and more prone to collective nonaffine deformations\cite{Lieleg2007,Huisman2011}. 

\begin{figure}
\begin{center}
\includegraphics[width=0.9 \columnwidth]{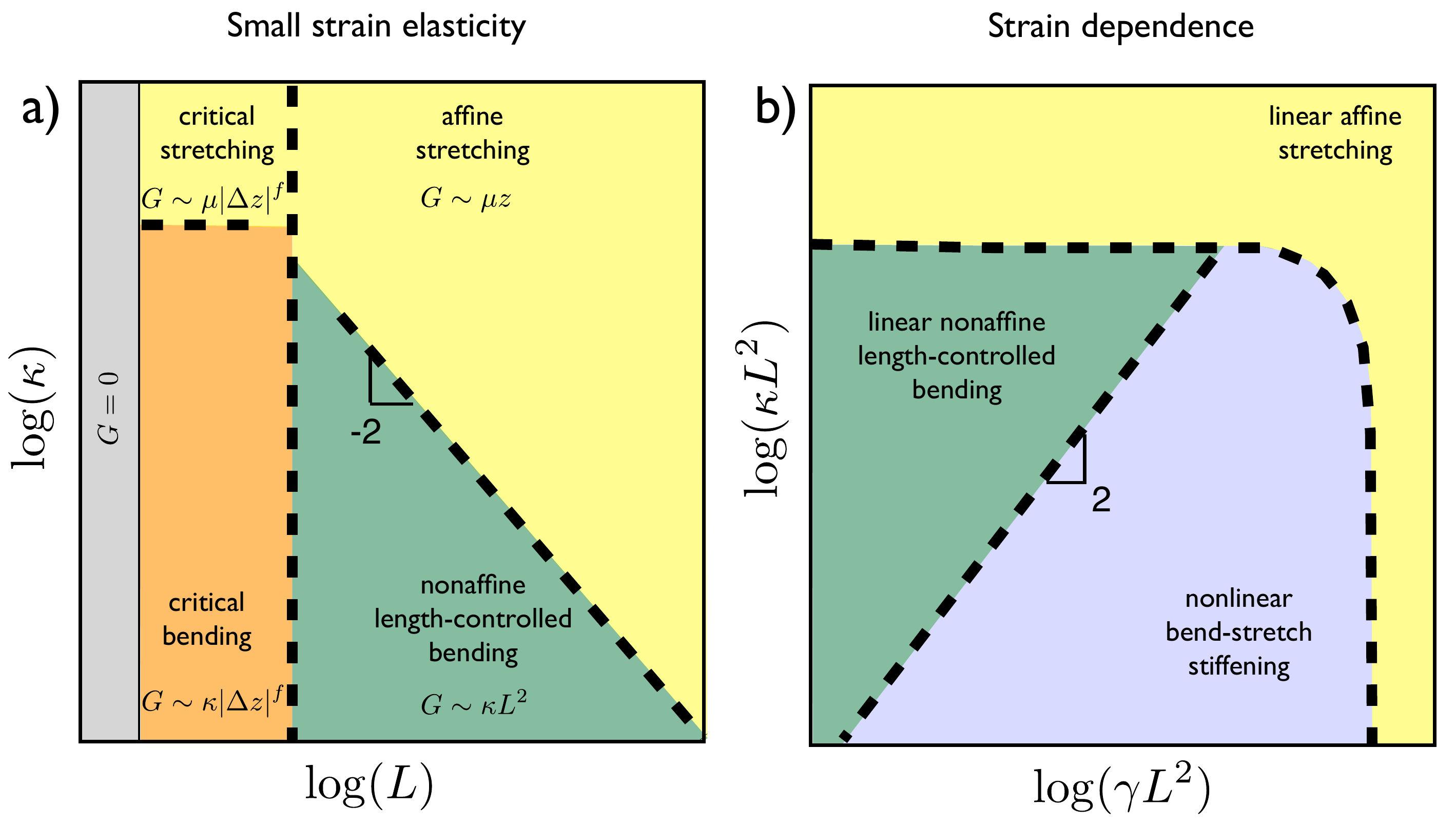}\vspace{-0.15in}
\caption{\label{fig:phase} (Color online) Phase diagrams for the linear (a) and nonlinear elasticity (b) of 3D fiber networks on the Phantom FCC lattice, where $L$ is the average filament length, $z$ is network connectivity, $\gamma$ is strain and $\kappa$ is the fiber bending rigidity. All lengths are measured in units of the lattice spacing $\ell_0$ and $\kappa$ in units of $\mu\ell_0^2$.}
\end{center}\vspace{-0.3in}
\end{figure}

Here we develop a numerical model for the elasticity of random 3D fiber networks with binary cross-links. This model provides access to a wide range of network densities below the CF isostatic threshold, as well as the previously inaccessible high-molecular weight limit. These networks exhibit various qualitatively distinct linear elastic regimes: a critical regime governed by the rigidity percolation point, a length-controlled bending regime and an affine stretching regime, as illustrated in Fig.~\ref{fig:phase}a. We provide a scaling analysis for insight into the origins of these regimes.  Network mechanics is determined by the ratio between a nonaffinity length scale and molecular weight. The high-molecular weight limit exhibits behavior reminiscent of zero-temperature critical behavior, including divergent strain fluctuations; these ultimately govern a crossover from the length-controlled bending regime to an affine, stretching dominated network response. This is remarkable, since the network remains well below Maxwell's CF isostatic connectivity threshold in this limit. Thus, paradoxically, although such networks can only be rigid at non-zero fiber bending stiffness, we find that no matter how weak this bending rigidity is, the network elasticity becomes insensitive to fiber bending in the limit of high---yet finite---molecular weight. Moreover, in the limit of floppy filaments with weak bending rigidity or high molecular weight, these networks become intrinsically nonlinear with a vanishing linear response regime (Fig.~\ref{fig:phase}b).

Much has been learned about stiff polymer gels from minimal models, such as 2D Mikado networks of randomly placed straight filaments with binary cross-links\cite{Head2003,Wilhelm2003}. The elasticity of such Mikado networks is governed by \emph{nonaffine} fiber bending deformations at low densities, while higher density networks exhibit predominantly \emph{affine} stretching elasticity of single fiber segments. This nonaffine-affine (NA-A) transition can be understood as being the result of increasing fiber-length. However, for such 2D networks, this high molecular weight limit actually coincides with Maxwell's CF isostatic connectivity, $\zcf=2d$ in $d$ dimensions\cite{Maxwell1864}, which can also give rise to a NA-A transition\cite{Broedersz2011}; it is thus unclear whether the observed transition in 2D is controlled by CF stretching constraints, or by filament length, as previously suggested by scaling arguments and floppy mode theory\cite{Head2003,Wilhelm2003,Heussinger2006}. However, 3D networks with binary cross-links---like most biopolymer systems---are qualitatively different; in this case the high-molecular weight limit corresponds to network connectivities well below $\zcf$. In the absence of fiber bending resistance, such networks exhibit zero-energy deformation modes and hence, they do not resist shear stresses. Thus, there are reasons to question the existence of an affine limit in realistic 3D networks with fibers that are softer to bending than to stretching\cite{Huisman2011,Heussinger2006,Buxton}. This is still subject of debate since studies in 3D have so far been limited to small systems\cite{Huisman2011} or to networks with high connectivities\cite{Buxton,Broedersz2011}. 

\begin{figure}
\begin{center}
\includegraphics[width=0.9 \columnwidth]{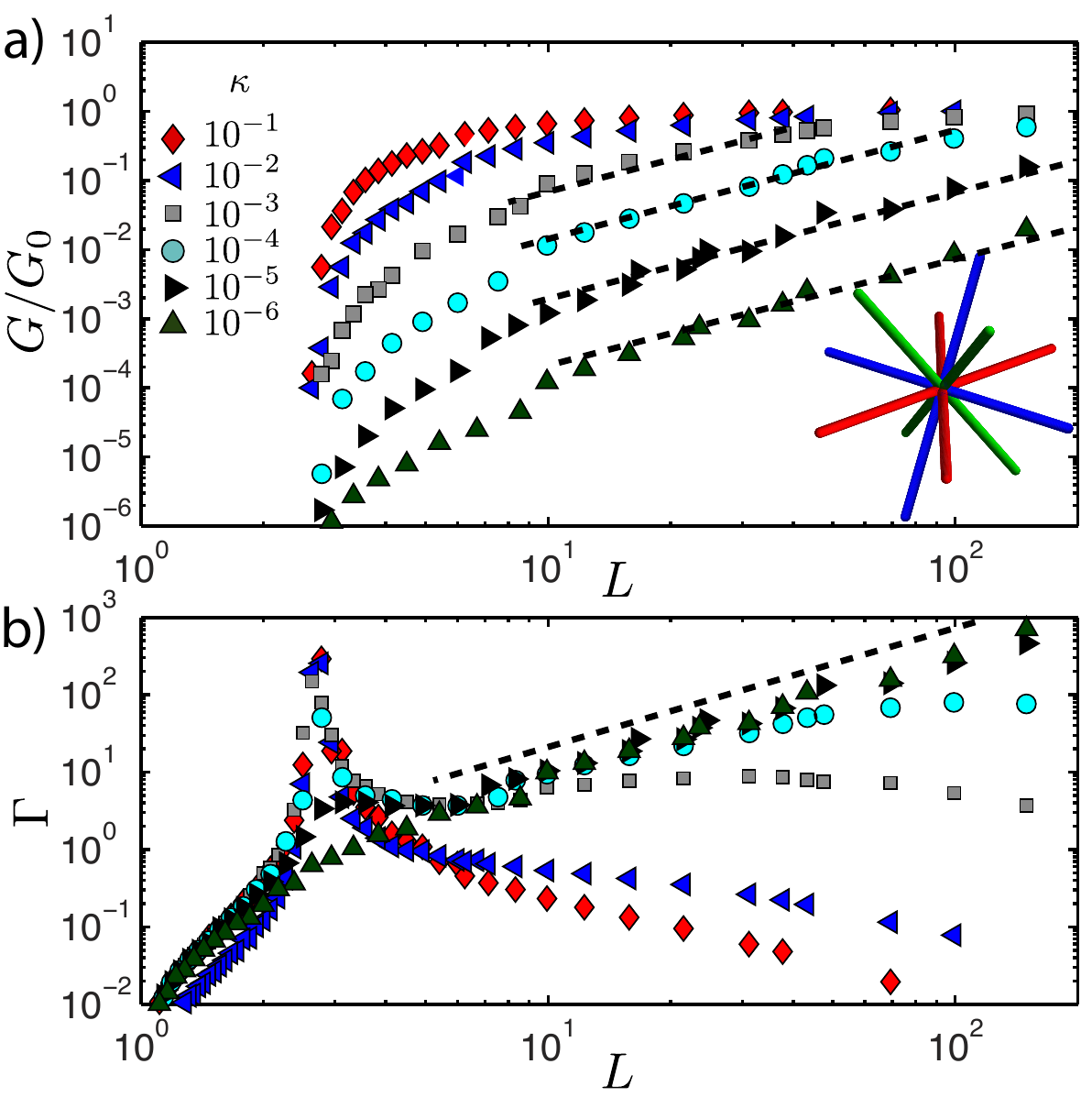}\vspace{-0.15in}
\caption{\label{fig:fig2} (Color online) a) The Shear modulus as a function of $L$ in units of $\ell_0$ for various $\kappa$ in units of $\mu\ell_0^2$. $G_0$ is the (affine) shear modulus of the undiluted Ph-FCC lattice. The inset illustrates the phantom principle: At each lattice vertex 3 independent binary cross-links are formed between randomly chosen fiber pairs  labeled by color. b) Non-affinity parameter $\Gamma$ as a function of $L$. Dashed black lines indicate a slope of 2.}
\end{center}\vspace{-0.3in}

\end{figure}

To provide insight in network mechanics at densities ranging from the rigidity percolation point to the high-molecular weight limit, we develop a 3D lattice-based fiber network model with binary cross-links. Our network's consists of straight fibers organized geometrically on a face centered cubic (FCC) lattice. However, we limit the maximum coordination number to four by randomly choosing three \emph{independent} pairs of cross-linked fibers out of the six fibers crossing at every vertex. Although the different binary cross-links may overlap geometrically, they do not constrain each other\cite{Broedersz2011a} (inset Fig.~\ref{fig:fig2}a). Therefore, we term this the Phantom FCC (Ph-FCC) lattice. This model is similar to a generalized Kagome lattice in 3D\cite{Stennul}, although it has a lower symmetry than the Ph-FCC lattice. By randomly cutting bonds with a probability, $1-p$, we tune the average molecular weight, $L=\ell_0 /(1-p)$, where $\ell_0$ is the distance between vertices\cite{Broedersz2011,Broedersz2011a}. 

The elastic energy of the 3D Ph-FCC network involves stretching and bending contributions of the constituent fibers, characterized by their stretching modulus $\mu$ and bending rigidity
$\kappa$. Each lattice vertex consists of 3 independent freely-hinging binary cross-links ranked by $h$. For small displacements, denoted by ${\bf u}_i^h$, the stretching energy of the network is expressed as
\begin{equation}
\label{eq:stretchenergy} E_{\rm S}=\frac{1}{2}\frac{\mu}{\ell_0}
\sum_{h=1}^3 \sum_{\langle ij \rangle} g_{ij}^h \left({\bf u}_{ij}^h \cdot {\bf \hat{r}}_{ij}  \right)^2,
\end{equation}
where the second sum extends over neighboring pairs of vertices, ${\bf u}_{ij}^h={\bf u}_{j}^h-{\bf u}_{i}^h$  and ${\bf \hat{r}}_{ij}$ is the bond direction in the undeformed lattice. Bond-dilution is implemented by setting $g_{ij}^h=1$ for present bonds and $g_{ij}^h=0$ for removed
bonds. Fibers form straight chains that resist angular deflections, leading to a total bending energy\cite{Das2007,Broedersz2011},
\begin{equation}
\label{eq:bendenergy}
E_{\rm B}=\frac{1}{2}\frac{\kappa}{\ell_0^{3}}
\sum_{h=1}^3 \sum_{\langle ijk \rangle} g_{ij}^h g_{jk}^h
\left[\left( {\bf u}_{ij}^h-{\bf u}_{jk}^h\right)\times {\bf \hat{r}}_{ij}\right]^2.
\end{equation}
Since the cross-links themselves do not provide a
torsional stiffness, the second sum only extends over \emph{coaxial} nearest neighbor triplets along the same fiber. The shear modulus, $G$, is determined numerically by applying a shear strain along the $111$-plane with Lees-Edwards periodic boundary conditions and energy minimizations are performed using a conjugate gradient algorithm.

Crucially, to avoid effects due to filaments spanning the network, thereby making unphysical stretch contributions to the elasticity of the sample, at least one bond is removed along 
every fiber. Our network sizes range from $W^3=20^3$ to $150^3$ unit cells, with up to three times that many cross-links. Due to the finite system size, this model can only approach $z=4$ asymptotically from below. 

\emph{Linear regime} We find that these networks have a finite shear rigidity only if $\kappa>0$, even though the perfect ($z=4$) Ph-FCC lattice  deforms affinely and has a finite shear modulus for $\kappa=0$. Nonetheless, over a broad range of $L$, the system can be bending dominated $G\sim\kappa$ (low $\kappa$), or stretching dominated, $G\sim\mu$ (high $\kappa$), as shown in Fig.~\ref{fig:fig2}a. Interestingly, there appear to be two distinct regimes well above the rigidity percolation point: a bending-dominated regime where $G$ depends strongly on $L$ (low $\kappa$ and $L$) and an $L$- and $\kappa$-independent affine stretching regime (high $\kappa$ and $L$).

\begin{figure}
\begin{center}
\includegraphics[width=0.9 \columnwidth]{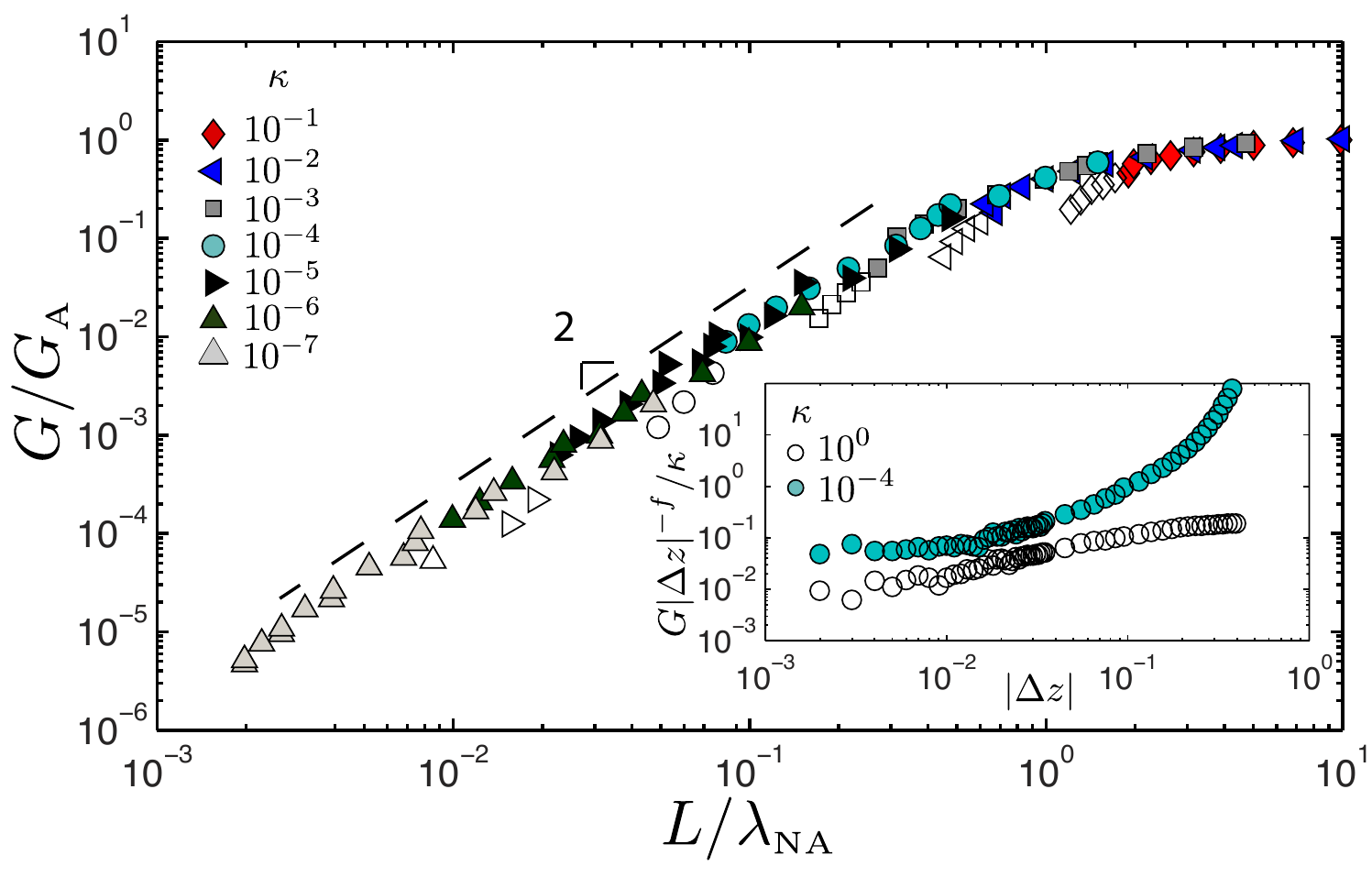}\vspace{-0.15in}
\caption{\label{fig:Hscale} (Color online) The shear modulus scaled with its affine prediction versus $L$ scaled with
$\lambda_{\mbox{\rm \tiny NA }} = \ell_0^2/\ell_b$ for various values of $\kappa$ in units of $\mu\ell_0^2$. The open symbols indicate data ranges in the rigidity percolation regime where we observe different scaling. The inset shows $G$ scaled with $|\Delta z|^{f}$ as a function of $|\Delta z|$.}
\end{center}\vspace{-0.3in}
\end{figure}

These results can be understood as follows. 
In the high-$\kappa$ limit, the system deforms affinely, with a shear modulus $G_{\mbox{{\rm \tiny A}}}\sim \frac{\mu}{\ell_0^2} z$. However, in the critical regime---controlled by the bending rigidity percolation point $z_b$---$G$ vanishes continuously with $\Delta z=z-\zb$\cite{Broedersz2011,Thorpe1983,Stennul,Wilhelm2003} as 
\begin{equation}
G_{\rm cs}\sim  \frac{\mu}{\ell_0^2}|\Delta z|^{f}, \ \  \ G_{\rm cb}\sim\frac{\kappa}{\ell_0^4} |\Delta z|^{f},
\end{equation}
for high and low $\kappa$, respectively. We find $z_b\approx2.4$ and $f\approx 0.65$ for a system size $W^3=30^3$, as demonstrated in the inset in Fig.~\ref{fig:Hscale} by showing that $G|\Delta z|^{-f}/\kappa$ reaches a plateau for low values of $\Delta z$. The rigidity threshold is similar to observations in prior 3D models\cite{Huisman2011}, although $f$ is considerably lower here, which is more consistent with findings on the generalized 3D Kagome lattice\cite{Stennul}. The rigidity threshold can be estimated using a Maxwell counting argument\cite{Maxwell1864,Huisman2011,Broedersz2011}; this connectivity-threshold occurs when per cross-link the number of  stretching constraints, $n_b z/4$, and bending constraints, $n_b (d-1)z^2/16$, equal the number of internal degrees of freedom $d$. Here, the number of bonds per cross-link $n_b=2$ in the undiluted network ($z=4$). This yields $z_b\approx2.6$, in reasonable agreement with the numerical results.

Since the CF isostatic point lies far beyond the physical connectivity range of this model, the naive expectation would be that this percolation regime extends over the whole range $z<4$. This would imply bending dominated network elasticity for low $\kappa$, such that $G\ll G_{\rm A}$ as $z\rightarrow4$ from below, with a discontinuous transition at $z=4$ to affine stretching dominated behavior due to fibers that span the whole network. However, this argument ignores possible effects due to filament length. In networks of straight fibers with binary interactions, the average fiber length diverges as $z \rightarrow4$ and large $L$ may lead to nonaffine displacements over greater length scales\cite{Heussinger2006}. The effects of high $L$ on the deformation field have been discussed in the context of 2D Mikado networks using both scaling arguments\cite{Head2003,Wilhelm2003} and floppy mode theory\cite{Heussinger2006}.

Here, we investigate the effects of molecular weight on the nonaffine deformation field and their implications for the mechanics of 3D fiber networks.
Network nodes along a fiber can undergo independent nonaffine deformations scaling as $\gamma L$ to avoid stretching of the other fibers to which they are connected. This direct scaling of nonaffine displacements with $L$ constitutes one of the central assumptions of floppy mode theory that was applied to Mikado networks\cite{Heussinger2006}.
To test this assumption here, we investigate the strain fluctuations using the nonaffinity measure\cite{Head2003,DiDonna2005,Broedersz2011}, 
$\Gamma=\frac{1}{\ell_0^2 \gamma^2} \big\langle (\delta {\bf u}^{\mbox{\rm \tiny NA}})^2\big\rangle$,
where $\delta {\bf u}^{\mbox{\rm \tiny NA}}={\textbf u}-\textbf{u}^{\mbox{\rm \tiny A}}$ denotes the nonaffine displacement of a cross-link and the brackets represent a network average. This nonaffinity measure exhibits a cusp at the bending rigidity percolation point, reflecting the criticality of the network's mechanics in this regime\cite{Broedersz2011,Wyart2008a}, as shown in Fig.~\ref{fig:fig2}b. Furthermore, there appears to be a regime for sufficiently low $\kappa$ where $\Gamma \sim L^2$ independent $\kappa$, lending credence to the basic assumption that $\delta u^{\mbox{\rm \tiny NA}}\sim L \gamma$\cite{Heussinger2006}.

Such nonaffine deformations imply a bending regime, 
\begin{equation}\label{Eq:GLC}
G_{\rm LC}\sim \frac{\kappa}{\ell_0^2} \left(\frac{\delta u^{\mbox{\rm \tiny NA}}}{\ell_0^2}\right)^2 \sim\frac{\kappa}{\ell_0^6} L^{2}.
\end{equation}
This prediction for the $L$-dependence of $G$ is in fact born out by the numerical data, as shown in Fig.~\ref{fig:fig2}a. This has the important implication that as $L$ becomes large, $G_{\rm LC}$ will exceed the affine shear modulus, and hence, an affine network deformation becomes more favorable. Thus, a NA-A transition\cite{Head2003,Das2007,Missel2010} occurs when $G_{\rm LC}\simeq G_{\rm A}$; this is satisfied when $L$ becomes comparable to the length scale
\begin{equation}
\lambda_{\mbox{\rm \tiny NA }} = \ell_0^2/\ell_b,
\end{equation}
where $\ell_b=\sqrt{\kappa/\mu}$.
Indeed, by plotting $G/G_{\rm A}$ as a function of $L/\lambda_{\mbox{\rm \tiny NA }}$ we find a good collapse of the data to a universal curve, for which $G/G_{\rm A}\approx1$ when $L/\lambda_{\mbox{\rm \tiny NA }}\gtrsim1$, as shown in Fig.~\ref{fig:Hscale}. This strongly supports the existence of a NA-A transition driven by molecular weight in 3D fiber networks with connectivities well below Maxwell's CF isostatic point.
In contrast, prior results for 2D networks suggested $\lambda_{\mbox{\rm \tiny NA }}\sim\ell_b^{-\alpha}$, with $\alpha\approx 0.3-0.4$\cite{Head2003,Wilhelm2003}; however, for such models it is unclear whether the NA-A transition is actually driven by fiber length, as for the 3D case presented here, or by the CF isostatic point\cite{Broedersz2011} that coincides with the high-$L$ limit for the Mikado model.

\begin{figure}
\begin{center}
\includegraphics[width= 0.9\columnwidth]{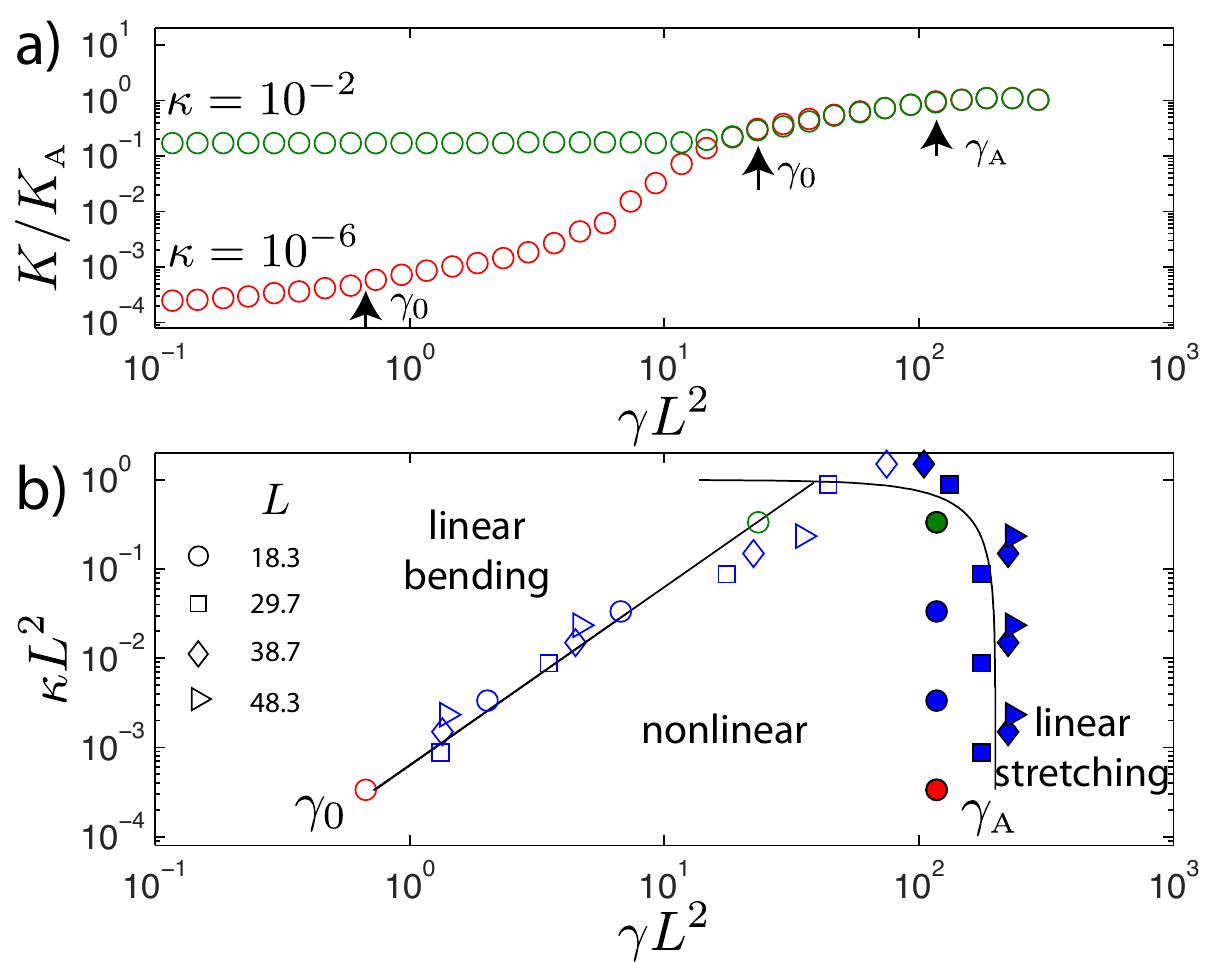}\vspace{-0.1in}
\caption{\label{fig:critstrain} (Color online) a) The differential modulus scaled with its affine prediction together with $\gamma_0$ and $\gamma_{\mbox{\rm \tiny A}}$ for networks with $L=18.3$. b) Scaling predictions (solid lines) and numerical data for $\gamma_0$ (open symbols) and $\gamma_{\mbox{\rm \tiny A}}$ (closed symbols). $L$ is measured in units of $\ell_0$ and $\kappa$ in units of $\mu\ell_0^2$. The red and green points in the lower panel correspond to $\kappa=10^{-6}$ and $\kappa=10^{-2}$, respectively.}
\end{center}\vspace{-0.3in}
\end{figure}

\emph{nonlinear regime} The length-controlled bending mechanics also has important implications for the nonlinear elasticity of 3D fiber networks. Even in a bending dominated regime, stretching modes are excited at finite network deformations\cite{Onck2005}, but to a higher order in the applied strain\cite{Heussinger2006,Lieleg2007,Wyart2008a}. Specifically, assuming length-controlled nonaffine deformations, a transverse bend with an amplitude $\sim\gamma L$ results in a stretch energy in the associated bond, $\delta E_S \sim \mu \epsilon^2 \ell_0$, where $\epsilon\sim(\gamma L/\ell_0)^2+O(\gamma^4)$. The onset of nonlinear network elasticity due to this effect occurs at a strain $\gamma_0$, at which $\delta E_S$ becomes comparable to the leading order bending contribution, $\delta E_B \sim \kappa L^2 \gamma^2/\ell_c^3 $. This stiffening saturates at a strain $\gamma_{\mbox{\rm \tiny A}}$, set by the condition $\delta E_B+\delta E_S\sim\frac{\mu}{\ell_0^2}\gamma^2$,  at which the network's response becomes affine. Thus, the onset and completion of the stiffening regime are expected to scale as 
\begin{equation}
\gamma_0\sim \frac{\ell_b}{L}, \ \ {\rm{and}} \ \  \gamma_{\mbox{\rm \tiny A}}\sim\frac{\ell_0^2}{L^2}\sqrt{1-L^2 \ell_b^2/\ell_0^4}.
\end{equation}
These predictions are in excellent agreement with the numerical results over a broad range of  $L$ and $\kappa$, as shown in Fig.~\ref{fig:critstrain}.
Interestingly, the nonlinear threshold $\gamma_0$ corresponds to a stress $\sigma_0\sim\kappa^{3/2} L$, which is distinct from the Euler buckling threshold~\cite{Onck2005}  and the stiffening threshold in sub-isostatic spring networks~\cite{Wyart2008a,Sheinman2011}.

The Phantom FCC model developed here, provides a powerful numerical model to probe the mechanics of 3D fiber networks with large system sizes. Using this model together with a scaling analysis, we have shown that even though the mechanical stability of 3D networks relies on the bending resistance of the constituent fibers, surprisingly for any $\kappa>0$, network mechanics becomes affine and independent of $\kappa$ when $L>\lambda_{\mbox{\rm \tiny NA }}$\footnote{For thermal semiflexible polymers we expect $\lambda_{\mbox{\rm \tiny NA }} \approx \sqrt{90 \ell_0 \ell_p}$~\cite{Head2003, MacKintosh95}.}. This NA-A transition is induced because it becomes energetically unfavorable to accommodate length-controlled non-affine deformations\cite{Heussinger2006} at large molecular weights.  Below this NA-A transition, in the length-controlled bending regime, we expect $G\sim\rho^{13/5}$, for semiflexible polymers for which the cross-linking lengthscale is expected to scale with $\rho^{-2/5}$\cite{MacKintosh95}, or $G\sim\rho^{3}$ for stiff polymers, where $\rho$ is the polymer length-density. These predictions may account for a recent report of $G\sim\rho^{2.68}$ in collagen networks\cite{Vader}.

We conjecture that the results for the length-controlled regime also apply to models with additional interactions, other than fiber bending, that stabilize the network below the CF-threshold---including next-nearest neighbor interactions or bond-bending interactions for cross-links that fix a preferred bond-angle. Specifically, we expect that such networks exhibit an affine high-molecular weight limit even for arbitrarily weak additional interactions.

\begin{acknowledgments}
This research was supported by the National Science Foundation under Grant No. NSF PHY05-51164 and by FOM/NWO. It's a pleasure to acknowledge discussions with E. Frey, M. Das, C. Heussinger, O. Stenull and T. C. Lubensky. We also thank OS and TCL for sharing their manuscript (Ref.~\cite{Stennul}) prior to publication.
\end{acknowledgments}


\begin{thebibliography}{50}\frenchspacing

\bibitem{Fletcher2010} D. A. Fletcher and R. D. Mullins, Nature {\bf 463}, 485 (2010).

\bibitem{Janmey1990} P. A. Janmey, S. Hvidt, J. Lamb, and T. P. Stossel, Nature {\bf 345}, 89 (1990).

\bibitem{Bausch2006} A. R. Bausch and K. Kroy, Nature Phys. {\bf 2}, 231 (2006).

\bibitem{Janmey2007} P. A. Janmey et al., Nature Materials, {\bf 6}: 48 (2007). 

\bibitem{Maxwell1864} J. C. Maxwell, Philos. Mag. {\bf 27}, 294 (1864).

\bibitem{Broedersz2011} C. P. Broedersz, X. Mao, T. C. Lubensky and F.C. MacKintosh. arXiv:1011.6535 (2011).

\bibitem{Wyart2008a} M. Wyart, H. Liang, A. Kabla and L. Mahadevan, Phys. Rev. Lett. {\bf 101}, 215501 (2008).

\bibitem{LiuNag1998} A. J. Liu and S. R. Nagel, Nature {\bf 396}, 21 (1998).

\bibitem{Thorpe1983} M. F. Thorpe, J. Non-Cryst. Solids {\bf 57}, 355 (1983).

\bibitem{Head2003} D. A. Head, A. J. Levine and F. C. MacKintosh, Phys. Rev. Lett. {\bf 91}, 108102 (2003); Phys. Rev. E {\bf 68}, 061907 (2003).

\bibitem{Wilhelm2003} J. Wilhelm and E. Frey, Phys. Rev. Lett. {\bf 91}, 108103 (2003).

\bibitem{Heussinger2006}  C. Heussinger and E. Frey, Phys. Rev. Lett. {\bf 97}, 105501 (2006);
C. Heussinger, B. Shaefer and E. Frey, Phys. Rev. E {\bf 76}, 031906 (2007).

\bibitem{Onck2005} P. R. Onck, T. Koeman, T. van Dillen and E. van der Giessen, Phys. Rev. Lett. {\bf 95}, 178102 (2005).

\bibitem{Das2007} M. Das, F. C. MacKintosh and A. J. Levine, Phys. Rev. Lett. {\bf 99}, 038101 (2007).

\bibitem{Missel2010} A. R. Missel M. Bai, W. S. Klug and A. J. Levine, Phys. Rev. E {\bf 82}, 041907 (2010).

\bibitem{Lieleg2007} O. Lieleg et al., Phys. Rev. Lett. {\bf 99}, 088102 (2007).

\bibitem{Vader} S. B. Lindstr{\"o}m, D. A. Vader, A. Kulachenko, and D. A. Weitz. Phys. Rev. E, {\bf 82}, 051905 (2010).

\bibitem{Huisman2011} E. M. Huisman, T. van Dillen, P. R. Onck, and E. Van der Giessen, Phys. Rev. Lett. {\bf 99}, 208103 (2007); E. M. Huisman  and T. C. Lubensky, Phys. Rev. Lett. \textbf{106}, 088301 (2011).

\bibitem{Buxton} G. A. Buxton and N. Clarke, Phys. Rev. Lett. {\bf 98}, 238103 (2007).
\bibitem{Stennul} O. Stenull and T. C. Lubensky, (to appear).

\bibitem{Broedersz2011a} C. P. Broedersz and F. C. MacKintosh, Soft Mater {\bf 7}, 3186 (2011).

\bibitem{DiDonna2005} B. A. Didonna and T. C. Lubensky, Phys. Rev. E {\bf72}, 066619 (2005).

\bibitem{Sheinman2011} M. Sheinman, C. P. Broedersz and F. C. MacKintosh (to appear).

\bibitem{MacKintosh95} F. C. MacKintosh, J. K\"{a}s and P. Janmey
Phys. Rev. Lett. {\bf 75}, 4425 (1995). 



\end{thebibliography}
\end{document}